\documentclass[a4paper,UKenglish,cleveref, autoref, thm-restate,authorcolumns]{lipics-v2019}
\usepackage{amsmath}
\usepackage{amssymb}
\usepackage{amsthm}
\usepackage{todonotes}
\usepackage{algorithm}
\usepackage{algpseudocode}
\usepackage{enumerate}
\usepackage{tikz}
\usepackage{hyperref}
\usepackage{lineno}

\usepackage{thmtools}
\usepackage{mathtools}

\usetikzlibrary{arrows.meta}
\usepackage[symbol]{footmisc}

\title{A sub-quadratic algorithm for the longest common increasing subsequence problem}

\author{Lech Duraj}{Theoretical Computer Science, Faculty of Mathematics and Computer Science, Jagiellonian University, Kraków, Poland}{duraj@tcs.uj.edu.pl}{https://orcid.org/0000-0002-0004-3751}{}

\authorrunning{L. Duraj}

\Copyright{Lech Duraj}



\funding{This work was partially supported by Polish National Science Center grant 2016/21/B/ST6/02165.}

\acknowledgements{I would like to thank Grzegorz Guśpiel, Grzegorz Herman and Adam Polak for helpful discussions and proofreading the paper, as well as the reviewers for many insightful comments and suggestions.}

\nolinenumbers 

\hideLIPIcs  

\EventEditors{Christophe Paul and Markus Bl\"{a}ser}
\EventNoEds{2}
\EventLongTitle{37th International Symposium on Theoretical Aspects of Computer Science (STACS 2020)}
\EventShortTitle{STACS 2020}
\EventAcronym{STACS}
\EventYear{2020}
\EventDate{March 10--13, 2020}
\EventLocation{Montpellier, France}
\EventLogo{}
\SeriesVolume{154}
\ArticleNo{37}

\newcommand{\fun}[2]{\textsc{#1}(#2)}

\newcommand{\lcis}[1]{\mathop{lcis}(#1)}

\newcommand{\lcisto}[1]{\mathop{lcis}^{\to}(#1)}
\newcommand{\prv}[1]{\pi(#1)}
\newcommand{\prev}[2]{\pi^{#1}(#2)}

\newcommand{\col}[1]{\mathop{symbol}(#1)}

\newcommand{\lo}[1]{\mathop{lo}(#1)}
\newcommand{\hi}[1]{\mathop{hi}(#1)}

\newcommand{\fullversion}[1]{#1}  
\newcommand{\shortversion}[1]{}  

\let\O\undefined
\newcommand{\O}[1]{\mathcal{O}\left(#1\right)}
\let\littleo\undefined
\newcommand{\littleo}[1]{o\left(#1\right)}
\newcommand{\Om}[1]{\Omega\left(#1\right)}
\let \Th\undefined
\newcommand{\Th}[1]{\Theta\left(#1\right)}

\begin{document}

\maketitle

\begin{abstract}
The Longest Common Increasing Subsequence problem (LCIS) is a natural variant of the celebrated Longest Common Subsequence (LCS) problem. For LCIS, as well as for LCS, there is an $\O{n^2}$-time algorithm and a SETH-based conditional lower bound of $\O{n^{2-\varepsilon}}$. For LCS, there is also the Masek-Paterson $\O{n^2 / \log{n}}$-time algorithm, which does not seem to adapt to LCIS in any obvious way. Hence, a natural question arises: does any (slightly) sub-quadratic algorithm exist for the Longest Common Increasing Subsequence problem?
We answer this question positively, presenting a $\O{n^2 / \log^a{n}}$-time algorithm for $a = \frac{1}{6}-\littleo{1}$. The algorithm is not based on memorizing small chunks of data (often used for logarithmic speedups, including the ``Four Russians Trick'' in LCS), but rather utilizes a new technique, bounding the number of significant symbol matches between the two sequences. 
\end{abstract}

\clearpage

\section{Introduction}

The Longest Common Increasing Subsequence problem (LCIS) is a variant of the well-known and extensively studied Longest Common Sequence (LCS) problem. The LCS is formulated as follows: given two integer sequences $A = (A[1], \ldots, A[n])$ and $B = (B[1], \ldots, B[n])$, determine another sequence $C$ which is a subsequence of both $A$ and $B$, of maximal possible length. In the LCIS variant, we require $C$ to be a strictly increasing subsequence. 

For LCS, a simple algorithm working in $\O{n^2}$-time was published in 1974 by Wagner and Fischer \cite{WagnerF74}. The complexity was later brought down to $\O{\frac{n^2}{\log{n}}}$ (for constant alphabet size) by Masek and Paterson, using a technique informally called the ``Four Russians trick'' \cite{MasekP80}. Some improvements have been made since then (in particular, \cite{Grabowski16} shaves another logarithm, down to $\O{\frac{n^2 \log \log n}{\log^2{n}}}$ even with arbitrary alphabet size), but no truly sub-quadratic, $\O{n^{2-\epsilon}}$-time algorithm has been found. There is even substantial evidence that a better algorithm might in fact not exist: it was shown by Abboud, Backurs and Vassilevska-Williams \cite{AbboudBVW15}, as well as by Bringmann and K\"{u}nnemann \cite{BringmannK18} that a truly sub-quadratic algorithm for LCS would yield a $2^{\delta n}$-time algorithm for SAT, with some $\delta < 1$, thus refuting the Strong Exponential Time Hypothesis (which states, roughly speaking, that such an algorithm is impossible \cite{ImpagliazzoP01, ImpagliazzoPZ01}). Hence, if we believe that SETH is true, then we must accept that no fast algorithms for LCS will ever be found. It is worth noting that in recent years several other SETH-based quadratic-time bounds were also shown, e.g., \cite{BackursI15} and \cite{RodittyVW13}.

As for LCIS, it is arguably one of the most interesting variants of LCS: neither of these problems seems to be reducible to the other (unless we count the reduction of LCIS to $3$-sequence LCS \cite{JacobsonV92}, which does not seem strong enough to have meaningful consequences). Therefore no algorithm or hardness result for LCS can be easily translated to a corresponding result for LCIS. The ``obvious'' dynamic programming algorithm for LCIS is $\O{n^3}$, the first $\O{n^2}$-time algorithm was given in \cite{YangHC05}, and possibly the simplest one was explicitly stated in \cite{ZhuW17}. A conditional lower bound was proven in \cite{DurajKP17}: it turns out that, as for LCS, any $\O{n^{2-\varepsilon}}$-time algorithm for LCIS would refute the Strong Exponential Time Hypothesis. The proof is based, like the one in \cite{AbboudBVW15}, on a reduction from the Orthogonal Vectors problem (introduced in \cite{Williams05}), but the reduction itself needs a quite different gadget construction. It is also worth mentioning that the problem of Longest Common Weakly Increasing Subsequence, similar to LCIS but with only weak monotonicity required, also has a conditional quadratic lower bound \cite{Polak17}. LCWIS, unlike LCIS, is also non-trivial when restricted to constant-size alphabets \cite{KutzBKK11, Duraj13}. It still remains an open question whether LCWIS admits a sub-quadratic algorithm for any alphabet size greater than $3$.

The LCIS problem itself has been studied quite extensively, and other algorithms have been proposed: Sakai \cite{Sakai06} found an algorithm which can retrieve the LCIS in linear space, Kutz et al.~\cite{KutzBKK11} presented an algorithm that works in (roughly speaking) $\O{n \cdot d}$ time, where $d$ is the output size (i.e. the length of LCIS). Chan et al.~\cite{ChanZFYZ07} proved that LCIS can be found in $\O{r \log \log n}$, where $r$ is the number of \emph{matching pairs} of symbols (i.e. the pairs $(x,y)$ with $A[x] = B[y]$). These algorithms work much faster for some specific cases (for example, they are sub-quadratic for ``random'' inputs with a reasonable notion of ``randomness''), but no algorithm that achieves $\littleo{n^2}$ worst-case complexity has been given so far. Arguably, one of the reasons is that the ``Four Russians Trick'' does not seem to adapt to current dynamic-programming LCIS algorithms -- at least, not in any easy way. In light of known conditional lower bound of this problem, we can only hope for complexity similar to $\O{\frac{n^2}{\log^a n}}$ for some $a>0$, but achieving this would seem interesting enough. ``The Art of Shaving Logs'', as called by Timothy M. Chan \cite{Chan13}, has already been practised for a variety of problems \cite{BaranDP05, Chan18, Williams07, LarsenW17, HanT12, Williams14}, sometimes yielding surprising results -- for example, some remarkable consequences in circuit complexity \cite{AbboudHWW16, AbboudB18}. Therefore, it appears natural to ask the question: \emph{Is there any slightly sub-quadratic (i.e. $\littleo{n^2}$-time) algorithm for LCIS?} 

This paper gives a positive answer to this question, by presenting an $\O{\frac{n^2 (\log \log n)^2}{\log^{1/6}{n}}}$-time algorithm for LCIS. Our algorithm iterates over matching pairs of symbols (as the one in \cite{ChanZFYZ07}), but to achieve sub-quadratic time, a new ,,log-shaving'' technique is introduced: we do not try to precompute the results for small chunks of data, as in LCS algorithms. Instead, we choose a useful subset of matching pairs -- so-called \emph{significant pairs} -- prove that there are $\littleo{n^2}$ such pairs, and adapt the algorithm to exploit this fact.

\section{Basic notions and paper outline}

Let $A$ and $B$ be the input sequences -- for most of the paper, it is convenient to allow $A$ and $B$ to have different lengths. Later, for the final complexity results, we will assume $|A| = |B|$. We use array-like notation for elements of $A$ and $B$, i.e. $A = (A[1], A[2], \ldots)$, $B = (B[1], B[2], \ldots)$. We will refer to the elements of $A$ and $B$ as \emph{symbols}, remembering that the symbols are in fact integers, and thus can be compared with each other. Also, we may assume that all those integers are positive and not exceeding $\O{|A|+|B|}$ -- if not, we can rename all the elements to be in range $\{1, 2, \ldots, |A|+|B|\}$, while preserving their relative order.

\begin{definition}[Matching pair]
A pair of indices $(x,y)$ for some $1 \leq x \leq |A|$, $1 \leq y \leq |B|$ is a \emph{matching pair} if $A[x] = B[y]$. For $\sigma = A[x] = B[y]$, we can say that $(x,y)$ is a $\sigma$-matching pair, or simply a $\sigma$-pair. We also say that $\sigma$ is the symbol of $(x,y)$ and sometimes write $\sigma = \col{x,y}$.
\end{definition}

\begin{definition}[Orders on pairs]
Let $(x,y)$ and $(x',y')$ be matching pairs. 
\begin{enumerate}[(1)]
\item We say that $(x,y) \leq (x',y')$ if $x \leq x'$ and $y \leq y'$. 
\item We say that $(x,y) \prec (x',y')$ if $x < x'$, $y < y'$ and $\col{x,y} < \col{x',y'}$. 
\end{enumerate}
\end{definition}

\begin{definition}[Common increasing subsequence]
 A common increasing sequence of $A$ and $B$ is a sequence of matching pairs $(x_1, y_1), \ldots, (x_s, y_s)$ such that $(x_1, y_1) \prec \ldots \prec (x_s, y_s)$.
\end{definition}

Our main problem is to find the longest possible common increasing subsequence. Sometimes, we wish to consider only some prefixes of $A$ and $B$, for which we will need the following two definitions:

\begin{definition}
 For any $x \leq |A|$ and $y \leq |B|$, we define $\lcis{x,y}$ as the maximal possible length of a common increasing subsequence that ends with some $(x',y')$ with $x' \leq x$ and $y' \leq y$. In other words $\lcis{x,y}$ is the length of the longest common increasing subsequence of $A[1..x]$ and $B[1..y]$.
\end{definition}

\begin{definition}
 For any matching pair $(x,y)$, we define $\lcisto{x,y}$ as the maximal possible length of a common increasing subsequence that ends with $(x,y)$. 
\end{definition}
\begin{remark}
 The value of $\lcis{x,y}$ is equal to $\max_{(x',y') \leq (x,y)}{\lcisto{x',y'}}$. In particular, for any $(x',y') \leq (x,y)$ we have $\lcisto{x',y'} \leq \lcis{x,y}$.
\end{remark}

A sequence realizing $\lcisto{x,y}$ must have some pair $(x',y')$ as the next-to-last element (providing that $\lcisto{x,y} \geq 2)$. Clearly, $\lcisto{x',y'} = \lcisto{x,y} - 1$. We call such a pair the \emph{predecessor} of $(x,y)$. There may be multiple candidates for the predecessor, so we break the ties first by $y$, then by $x$. Formally:

\begin{definition}[Predecessor]
 For a matching pair $(x,y)$ the predecessor $\prv{x,y}$ is a matching pair $(x',y') \prec (x,y)$ such that:
 \begin{enumerate}[(1)]
  \item $\lcisto{x',y'} = \lcisto{x,y}-1$,
  \item $(x',y')$ has the minimal possible $y'$ of all pairs satisfying (1),
  \item $(x',y')$ has the minimal possible $x'$ of all pairs satisfying (1) and (2).
 \end{enumerate}
\end{definition}

An example is shown in Figure \ref{figure:example-pred} below:

\begin{figure}[H]
\centering

\def\nodecolors{{"gray", "green", "blue", "cyan", "red"}}
\def\valA{{1, 3, 5, 2, 5, 4, 5 }}
\def\valB{{1, 2, 5, 3, 5, 4, 5 }}
\begin{tikzpicture}[scale=.9, transform shape]
\tikzstyle{every node} = [circle]
\node(A) [font=\large\bf] at (-1,3) {A:};
\foreach \x [count=\n] in {0,...,6}{%
    \pgfmathparse{\valA[\x]};
    \edef\cwhat{\pgfmathresult};
    \pgfmathparse{\nodecolors[\cwhat-1]};
    \edef\ccol{\pgfmathresult};
    \node(A\x)[shape=rectangle,minimum size=0.8cm, font=\large, fill=\ccol!30] at (\x,3) {\cwhat};
}
\node(B) [font=\large\bf] at (-1,0) {B:};
\foreach \x [count=\n] in {0,...,6}{%
    \pgfmathparse{\valB[\x]};
    \edef\cwhat{\pgfmathresult};
    \pgfmathparse{\nodecolors[\cwhat-1]};
    \edef\ccol{\pgfmathresult};
    \node(B\x)[shape=rectangle,minimum size=0.8cm, font=\large, fill=\ccol!30] at (\x,0) {\cwhat};
}
\foreach \source/\target/\val in {0/0/1, 3/1/2, 5/5/3, 6/6/4 }{%
    \pgfmathparse{\valA[\source]};
    \edef\cwhat{\pgfmathresult};
    \pgfmathparse{\nodecolors[\cwhat-1]};
    \edef\ccol{\pgfmathresult};
    \draw [\ccol,thick] (A\source) -- node[draw=\ccol,pos=0.3,shape=rectangle,style=solid,fill=white,scale=0.8,font=\small\bf](EU\source\target){\val} node[pos=0.7,shape=circle,scale=0.5](ED\source\target){} (B\target) ;
}

\foreach \source/\target/\val in {1/3/2, 2/4/3, 4/4/3}{%
    \pgfmathparse{\valA[\source]};
    \edef\cwhat{\pgfmathresult};
    \pgfmathparse{\nodecolors[\cwhat-1]};
    \edef\ccol{\pgfmathresult};
    \draw [\ccol,thick] (A\source) -- node[pos=0.7,shape=circle,scale=0.5](EU\source\target){\val} node[draw=\ccol,pos=0.7,shape=rectangle,style=solid,fill=white,scale=0.8,font=\small\bf](ED\source\target){\val} (B\target) ;
}

\draw[-Latex,color=red] (EU66) to [out=140, in=20] (EU55);
\draw[-Latex,color=cyan] (EU55) to [out=140, in=10] (EU31);
\draw[-Latex,color=red] (ED44) to [out=110, in=-20] (EU31);
\draw[-Latex,color=green] (EU31) to [out=140, in=20] (EU00);
\draw[-Latex,color=red] (ED24) to [out=200, in=-20] (ED13);
\draw[-Latex,color=blue] (ED13) to [out=200, in=-20] (EU00);
\end{tikzpicture}
 \caption{An example: for two sequences $A = (1, 3, 5, 2, 5, 4, 5)$ and $B = (1, 2, 5, 3, 5, 4, 5)$ some matching pairs are shown. A pair $(x,y)$ is labeled with $\lcisto{x,y}$ and an arrow leads from $(x,y)$ to $\prv{x,y}$. Some pairs were omitted for clarity.}
\label{figure:example-pred}
\end{figure}
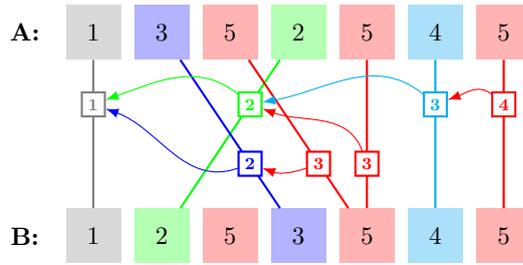

\begin{definition}
For a matching pair $(x,y)$ with $\lcisto{x,y} > k$, the $k$-th predecessor $\prev{k}{x,y}$ is defined inductively as $(x,y)$ for $k = 0$ and $\prv{\prev{k-1}{x,y}}$ for $k \geq 1$. In particular, $\prev{1}{x,y} = \prv{x,y}$.
\end{definition}

The algorithm for LCIS in \cite{ChanZFYZ07} iterates over all matching pairs in $A$ and $B$. There may be, however, as many as $\Theta(n^2)$ of them -- it is easy to construct an example of such sequences by including a lot of equal elements. Observe, though, that some of the matching pairs may not really matter in the solution -- for example, if $A[x+1] = A[x]$, then a matching pair $(x+1,y)$ for any $y$ is as good as $(x,y)$, and we could drop $A[x+1]$ from $A$ altogether. We generalize this observation to form the notion of a \emph{significant pair}, which is the central concept of this paper, allowing us to construct the desired faster algorithm for LCIS.

\begin{definition}[Significant pair]
 Let $(x,y)$ be a $\sigma$-pair, i.e. $\sigma = A[x] = B[y] $. We say that $(x,y)$ is a significant pair if for every $\sigma$-pair $(x',y') \leq (x,y)$, if $(x',y') \neq (x,y)$ then $\lcisto{x',y'} < \lcisto{x,y}$.
\end{definition}

Again, we include an example to make this important definition more clear:

\begin{figure}[H]
\centering
     
\def\nodecolors{{"gray", "green", "blue", "cyan", "red"}}
\def\valA{{1, 3, 5, 2, 5, 4, 5 }}
\def\valB{{1, 2, 5, 3, 5, 4, 5 }}
\begin{tikzpicture}[scale=.9, transform shape]
\tikzstyle{every node} = [circle]
\node(A) [font=\large\bf] at (-1,3) {A:};
\foreach \x [count=\n] in {0,...,6}{%
    \pgfmathparse{\valA[\x]};
    \edef\cwhat{\pgfmathresult};
    \pgfmathparse{\nodecolors[\cwhat-1]};
    \edef\ccol{\pgfmathresult};
    \node(A\x)[shape=rectangle,minimum size=0.8cm, font=\large, fill=\ccol!30] at (\x,3) {\cwhat};
}
\node(B) [font=\large\bf] at (-1,0) {B:};
\foreach \x [count=\n] in {0,...,6}{%
    \pgfmathparse{\valB[\x]};
    \edef\cwhat{\pgfmathresult};
    \pgfmathparse{\nodecolors[\cwhat-1]};
    \edef\ccol{\pgfmathresult};
    \node(B\x)[shape=rectangle,minimum size=0.8cm, font=\large, fill=\ccol!30] at (\x,0) {\cwhat};
}
\foreach \source/\target/\val in {2/2/2, 4/2/3, 2/4/3, 6/6/4 }{%
    \draw [red,thick] (A\source) -- node[draw=red, shape=rectangle, fill=white, pos=0.25, font=\small\bf] {\val} (B\target) ;
}
\foreach \source/\target/\val in {4/4/3, 4/6/3, 6/4/3}{%
    \draw [red,dashed] (A\source) -- node[draw=red, style=solid, shape=rectangle, fill=white, pos=0.25, font=\small] {\val} (B\target) ;
}

\end{tikzpicture}
 \caption{An example of two sequences $A$ and $B$ with some matching pairs. A pair $(x,y)$ is labeled with $\lcisto{x,y}$; the significant pairs are drawn with solid lines, while the insignificant ones -- with dashed lines. Some pairs were ommitted for clarity.}
\label{figure:example-significant}
\end{figure}
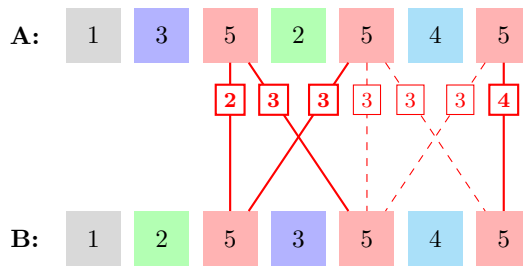

\begin{claim}
 If $(x,y)$ is a matching pair, then $\prv{x,y}$, if exists, is a significant pair.
 \begin{proof}
  Easy from the tie-breaking rule in the predecessor definition: let $(x',y') = \prv{x,y}$. If $(x',y')$ is not significant, then there is a better candidate for the predecessor.
 \end{proof}
\end{claim}

Having defined the significant pairs, we propose the following two theorems, which together form our main result. The first bounds the number of such pairs, the second proposes an algorithm that exploits this bound:

\begin{theorem}
\label{thm:main-count}
 For any $A$, $B$ with $|A|,|B| \leq n$, the number of significant pairs is at most $\O{\frac{ n^2}{\log^{1/3}{n}}}$.
\end{theorem}

\begin{theorem}
\label{thm:main-algo}
 Suppose that $|A|, |B| \leq n$ and that there are at most $\O{\frac{n^2}{t}}$ significant pairs, with $t = t(n)$ satisfying $\log t = \Th{\log \log n}$. There is an algorithm which finds LCIS in $\O{\frac{n^2 (\log \log n)^2}{\sqrt{t}}}$ time complexity.
\end{theorem}

The obvious consequence of Theorems \ref{thm:main-count} and \ref{thm:main-algo} is the following:

\begin{corollary}
There is an algorithm which finds LCIS of two sequences $A, B$ with $|A|, |B| \leq n$ in $\O{\frac{n^2 (\log \log n)^2}{\log^{1/6}{n}}}$ time complexity.
\end{corollary}

The rest of the paper is devoted to proving the two main theorems: in Section \ref{section:counting} we prove Theorem \ref{thm:main-count}, whereas Section \ref{section:algo-full} describes the algorithm of Theorem \ref{thm:main-algo}.

\section{Counting significant pairs}
\label{section:counting}

\subsection{The idea}
\label{section:idea}

In this section we present the high-level idea behind the bound for the number of significant pairs. Please note that while the proof originally stems from this concept, its final version needs also some careful counting and balancing arguments, as well as a few non-intuitive tricks. Therefore, we start with an informal sketch to give some intuitions, while the full, formal proof will be presented in Sections \ref{section:padding}-\ref{section:predmatrix}.

Imagine two sequences $A$ and $B$ with $|A| = |B| = n$ and with the number of significant pairs ,,very close'' to $\Th{n^2}$. This requires at least one symbol $\sigma$ generating a lot of significant $\sigma$-pairs itself (as opposed to, for example, $\sqrt{n}$ different symbols generating $\Th{n^{3/2}}$ pairs each -- this is impossible, as it would imply $|A| > n$ or $|B| > n$). We then focus on one particular such symbol $\sigma$ and imagine a graph $G_{\sigma}$ with all occurrences of $\sigma$ in $A$ and $B$ as vertices of $G_\sigma$, and all significant $\sigma$-pairs as its edges. (An example is already provided with Figure \ref{figure:example-significant} -- for $\sigma = 5$, the red elements of $A$ and $B$ are the vertices of $G_\sigma$, and red solid lines are the edges).

Denote by $s = s(n)$ the largest integer such that $G_\sigma$ has at least $s$ vertices of degree at least $s$ -- the total number of edges cannot exceed $\O{n \cdot s}$, so simplifying a little bit, our goal is to show that $s = \littleo{n}$. But every $\sigma$-pair must have its predecessors, which are $\tau$-pairs for some other symbols $\tau < \sigma$. The proof is based on the observation that these predecessors need quite a lot of different symbols  -- we argue that the total number of required elements of $A$ and $B$ is asymptotically greater than $s$. This forces $s = \littleo{n}$, and after careful calculations we obtain more specific bounds.

To take a closer (but still preliminary) look at our main tools, consider a vertex $A[x]$ in $G_\sigma$ with edges $(x,y_1), \ldots, (x,y_s)$ with $y_1 > y_2 > \ldots > y_s$. Consider, for a fixed $k > 0$, all predecessors $\prev{k}{x, y_i}$ for $1 \leq i \leq s$ (let us not worry, for the moment, whether the predecessors exist), denoting $\prev{k}{x, y_i} = (u_i, v_i)$. We claim that for every $i$, $v_{i+k} < v_{i}$. This is because $v_{i+k} < y_{i+k}$, while $v_{i} < y_{i+k}$ would lead to $(u_i,v_i) \prec (x,y_{i+k})$, which would in turn yield $\lcisto{x,y_{i+k}} \geq \lcisto{u_i,v_i}+1 = \lcisto{x,y_i}-k+1$. But the values $\lcisto{x,y_i}, \ldots, \lcisto{x,y_{i+k}}$ must all be different (as the pairs are significant), which implies $\lcisto{x,y_{i+k}} \leq \lcisto{x,y_{i}}-k$ -- a contradiction. Therefore $v_{i} > y_{i+k} > v_{i+k}$.
This shows that there must be at least $\frac{s}{k}$ different elements in $B$ to accommodate the $k$-th predecessors of the pairs $(x,y_i)$ for $i = 1, 2, \ldots, s$. A notable edge case is that for every $k$, all those predecessors happen to use the same symbol $\tau_k$, implying at least $\frac{s}{k}$ occurrences of $\tau_k$ in $B$ and thus at least $s + s/2 + s/3 + \ldots = \Om{s \log s}$ symbols in $B$, which would immediately imply $s = \O{n/\log{n}}$. Of course, we cannot hope to be that fortunate, but we can salvage some bounds from this argument: if, for some $\delta = \delta(n)$, which is $\omega(1)$ and $\littleo{\log n}$, and for any particular $x\in A$, the $k$-th predecessors use at most $\delta$ different symbols for every $k$, we can prove that this yields $\Om{\frac{s \log s}{\delta}}$ symbols in $B$, so $s = \O{ \frac{n \delta}{\log n}} = \littleo{n}$ and we are done -- this is formally proven in more general form in Section \ref{section:predmatrix}. On the other hand, if there are more than $\delta$ different symbols among predecessors for every possible $x$, we expect $\delta$ different elements before $x$ in $A$. Using similar arguments as before, we argue that those sets of elements must be (at least partly) disjoint for different picks of $x$. But there are $\Om{s}$ possible choices of a high-degree vertex $x$, which also implies $s = \littleo{n}$ -- this sketch roughly corresponds to Section \ref{section:sigmapairs}.

We now move on to the full proof. Before presenting the core observations, we start with padding the sequences -- adding dummy elements to make computing predecessors easier. 

\subsection{Preliminaries -- padding the sequences}
\label{section:padding}
Consider two sequences $A$, $B$ with $|A|, |B| \leq n$. Suppose that $\lcis{|A|,|B|} \geq 1$ -- if there are no common elements in $A$ and $B$, there is nothing to be proven. We can also assume $n \geq 2$. For the sake of analyzing significant pairs between $A$ and $B$, we shall modify the sequences a little bit. First, we can assume, without loss of generality, that $A$ and $B$ contain only positive integers (adding any constant to all the elements does not change anything). Then we pad both the sequences, inserting a prefix of dummy elements $P_n = (-2n, -2n+1, -2n+2, \ldots, -1)$ in the front, obtaining new sequences $\hat{A}$ and $\hat{B}$. More precisely, we put $\hat{A} = P_n \circ A$ and $\hat{B} = P_n \circ B$, where $\circ$ is the operator of sequence concatenation. These new elements now contribute to all previous common increasing subsequences, increasing their lengths by exactly $2n$. This does not change the significance of any ``old'' matching pairs, i.e. any significant pair $(x,y)$ present in $A$ and $B$ remains a significant pair $(2n+x, 2n+y)$ in $\hat{A}$ and $\hat{B}$. Therefore the number of significant pairs can only increase in this operation, so it is enough to prove the inequality of Theorem \ref{thm:main-count} for $\hat{A}$ and $\hat{B}$. The padding operation also ensures that for every matching pair $(x,y)$ with $x,y > 2n$ we have $\lcisto{x,y} > 2n$, allowing us to compute up to $2n$ predecessors of $(x,y)$.

\subsection{The $\sigma$-pair graph}
\label{section:sigmapairs}

First, let $\delta = \delta(n) =  \frac{\log^{1/3}{n}}{4}$ -- our goal is to bound the number of significant pairs by $\O{\frac{n^2}{\delta(n)}}$, and the order of magnitude of $\delta$ is chosen as ,,the highest one for which the proof still holds''.

The crucial step of the proof starts with fixing a symbol $\sigma$ and bounding only the number of significant $\sigma$-pairs, which we will then sum up over all possible $\sigma$. Without loss of generality we remove -- for a while -- all elements greater than $\sigma$ from $\hat{A}$ and $\hat{B}$, as they do not affect the significance of any $\sigma$-pair. 

Let $A_\sigma$ (resp.~$B_\sigma$) be the set of all positions $x$ with $\hat{A}[x] = \sigma$ (resp.~$\hat{B}[x] = \sigma)$. Consider a bipartite graph $G_\sigma$ with the set of vertices $V(G_\sigma) = A_\sigma \cup B_\sigma$, and the edge set $E(G_\sigma) = \{(x,y) \in A_\sigma \times B_\sigma : (x,y)\ \mbox{is a significant pair} \}$. Our main goal is to bound the number of edges in $G_\sigma$. To do that, we first define some family of ,,bad'' configurations of edges and prove that if any of them is forbidden, the number of edges can be bounded as desired. Finally -- which is the most technical part -- we show that at least one of those configurations does not, indeed, appear in the graph.

For any $1 \leq x \leq |\hat{A}|$, let us denote by $A_x^{\to}$ the suffix of $\hat{A}$ starting at $x$ (including $x$). For an integer $k$ we say that $A_x^{\to}$ is a \emph{$k$-dense suffix}, if:

\begin{itemize}
    \item $|A_x^{\to}| \leq \lceil k\cdot\delta \rceil$,
    \item There are $\lfloor n/\delta \rfloor$ distinct edges $(x,c_1), \ldots, (x,c_{\lfloor n/\delta \rfloor}) \in G_\sigma$,
    \item Every $c_i$ has, in turn, $k$ distinct edges $(y_{ij},c_i)$ for $j = 1, 2, \ldots, k$ and some $y_{ij} \in A_x^{\to}$.
\end{itemize}

The following lemma is the core idea of the proof, as it forbids at least one dense suffix to appear in $G_\sigma$. As its proof needs careful analysis (boiling down to counting predecessors of $\sigma$-pairs), and is somewhat technical, we defer the proof until Section \ref{section:predmatrix}. Before that, we will use Lemma \ref{lemma:no-bad} to show our ultimate goal, Theorem \ref{thm:main-count}.

\begin{restatable}{lemma}{LemmaNoBad}
\label{lemma:no-bad}
For every $A$ and $B$ with $|A|, |B| \leq n$, and for the corresponding graph $G_\sigma$, there exists some positive integer $k \leq n/\delta$ such that there is no $k$-dense suffix in $\hat{A}$.
\end{restatable}

 To prove that this lemma bounds the number of edges in $G_\sigma$, we first split vertices of $\hat{A}$ according to their degree. The \emph{small} vertices are these with degree at most $\frac{2n}{\delta}$, the rest being \emph{large} vertices. The following observation is straightforward:

\begin{remark}
\label{lemma:one-symbol-small}
 The total number of edges incident to small vertices of $\hat{A}$ is at most $|A_\sigma| \cdot \frac{2n}{\delta}$.
\end{remark}

It remains to bound the number of edges incident to large vertices in $\hat{A}$. For every connected substring $S \subseteq \hat{A}$, let us denote by $L(S)$ the number of such edges between $S$ and $\hat{B}$.

\begin{lemma}
\label{lemma:one-symbol-ineq}
For every $A$ and $B$ with $|A|, |B| \leq n$, $L(\hat{A}) \leq |B_\sigma| \cdot \frac{2|\hat{A}|}{\delta}$.
\begin{proof}

We use induction on $|\hat{A}|$ (please note that $n$ remains fixed throughout the proof, so $|\hat{A}|$ is always equal to $|A|+2n$). The minimal length of a padded sequence is $|\hat{A}| = 2n+1$ -- i.e. with only one non-dummy element -- and in this case we have $|E| \leq |B_\sigma|$. Suppose now that we have some $\hat{A}$ with $|\hat{A}| = a$, and have already proven the statement for all $A'$ with $|A'| < a$.

Let $k$ be the integer obtained from Lemma \ref{lemma:no-bad} and let $Y$ be the suffix of of $\hat{A}$ of length $\lceil k \delta \rceil$ (as $\lceil k \delta \rceil \leq n$, $Y$ is a proper suffix).  We can now use Lemma \ref{lemma:no-bad} to bound the number of edges between $Y$ and $\hat{B}$. Let us initially place $k$ tokens on every $\sigma$-element of $\hat{B}$ and consider all elements of $Y$, starting from the last one. For every large vertex $x \in Y$ we look at all its neighbors and remove one token from each of them, whenever they still have one. We claim that every time, at least half of these neighbours (which is at least $n/\delta$, as we are dealing with large vertices) must still have a token to spare. This is because if at least $n/\delta$ neighbors of $x$ were already tokenless, than $A_x^{\to}$ would be a $k$-dense suffix: clearly $|A_x^{\to}| \leq \lceil k\delta \rceil$, and we have just found that $x$ has $\lfloor n/\delta \rfloor$ neighbors which have already lost their $k$ tokens, so each of them has $k$ neighbours in $A_x^{\to}$. 

Therefore, every $x$ must be able to take a token from at least half of its neighbors. As there are only $k\cdot |B_\sigma|$ tokens to be removed, the total number of edges $L(Y)$ cannot exceed $2k |B_\sigma| \leq 2 \cdot \frac{\lceil k\delta \rceil }{\delta} \cdot |B_\sigma| = |Y| \cdot \frac{2|B_\sigma|}{\delta}$. 

Let us denote by $\hat{A}-Y$ the prefix of $\hat{A}$ obtained by deleting $Y$ from the end of $\hat{A}$ (i.e. the prefix of length $|\hat{A}| - |Y|$). Now if $|\hat{A}-Y| \leq 2n$ (i.e. $Y$ uses up all non-padding symbols), then $L(\hat{A}-Y) = 0$. Otherwise, we can apply the induction hypothesis to $\hat{A}-Y$, obtaining $L(\hat{A}-Y) \leq |\hat{A}-Y| \cdot \frac{2|B_\sigma|}{\delta}$. In both cases we have $L(\hat{A}) = L(\hat{A}-Y) + L(Y) \leq |\hat{A}|\cdot \frac{2|B_\sigma|}{\delta}$, as desired.

\end{proof}
\end{lemma}

We can now prove Theorem \ref{thm:main-count} and bound the total number of significant pairs:

 \begin{proof}[Proof of Theorem \ref{thm:main-count}]
 We want to show that for any $A$, $B$ with $|A|,|B| \leq n$, the number of significant pairs is at most $\O{\frac{n^2}{\log^{1/3}{n}}}$. We already know that is enough to prove it for padded sequences $\hat{A}$ and $\hat{B}$. For a fixed $\sigma$, the number of significant $\sigma$-pairs is, from Remark \ref{lemma:one-symbol-small} and Lemma \ref{lemma:one-symbol-ineq}, at most $|A_\sigma| \cdot \frac{2n}{\delta} + |B_\sigma| \cdot \frac{2|\hat{A}|}{\delta} \leq (|A_\sigma| + |B_\sigma|) \cdot \frac{6n}{\delta}$. Summing this over all possible symbols $\sigma$ (and using the fact that $\sum_\sigma (|A_\sigma|+|B_\sigma|) = |\hat{A}|+|\hat{B}| \leq 6n$), we get that the total number of significant pairs does not exceed $\frac{36 n^2}{\delta} \leq \frac{144 n^2}{\log^{1/3}{n}}$.
 \end{proof}

To close this section, it may be worth asking if our bound of $\O{\frac{n^2}{\log^{1/3}{n}}}$ significant pairs is tight, or at least close to the optimal one. We partially answer that in the \shortversion{full version of the paper}\fullversion{Appendix \ref{sec:howmanypairs}}, providing an example of two sequences with $\Om{\frac{n^2}{\log{n}}}$ significant pairs. Hence, we cannot go lower than this bound, but there is still a gap for possible future work.

\subsection{Dense suffixes and the predecessor matrix}
\label{section:predmatrix}

In this section we complete the missing part by proving Lemma \ref{lemma:no-bad}. To do that, we need to introduce a new concept -- the \emph{predecessor matrix} $M$ of significant pairs. To give some intuition what this matrix is, imagine that we first find the longest common increasing subsequence of $\hat{A}$ and $\hat{B}$ and put this sequence of significant pairs into the first column of $M$, one pair in every cell (starting from last element of LCIS, downwards). Then we delete some final elements of $\hat{B}$ such that the length of LCIS decreases by exactly 1, find the new (possibly very different) LCIS, and form $M$'s second column the same way. We can repeat this process $n$ times, and the padding of the sequences always allows us to compute $n$ predecessors. 

Each entry of $M$ is a significant pair $(x,y)$. We refer to $\col{x,y}$ as \emph{color} of this entry of $M$, as we feel this gives a better intuition. To analyze $M$, we look at the number of different colors in each row. We show that too few colors in every row would cause $\hat{B}$ to accumulate more than $3n$ elements -- which is impossible -- so there is a row (say, $k$-th) with somewhat more colors -- we then prove that this row corresponds to the desired $k$ fulfilling the statement of Lemma \ref{lemma:no-bad}.

To formally define $M$, recall the previous assumptions: we have sequences $\hat{A}$ and $\hat{B}$ which are both padded with $P_n$ and do not contain symbols greater than $\sigma$. Let $a = |\hat{A}|$, $b = |\hat{B}|$ and $\ell = \lcis{a,b}$. Because of padding we know that $2n < a, b \leq 3n$ and that $\ell > 2n$. It is easy to see that for every $y>1$ we have $\lcis{a,y-1} \geq \lcis{a,y}-1$. Therefore, if we iterate $y$ downwards from $b$ to $1$, $\lcis{a,y}$ takes all values between $\ell$ and $1$. In particular, there must exist elements $b = b_1 > b_2 > \ldots > b_n$ such that:

$$\lcis{a,b_j} = \ell - j + 1,$$

for every $j = 1, 2, \ldots, n$.

We define the \emph{predecessor matrix} $M$ as an $n \times n$ matrix of matching pairs. For $1 \leq j \leq n$ we consider the longest common increasing sequence realizing $\lcis{a,b_j}$ and define $M[1,j]$ as its last element $(x^*, y^*)$. If there are multiple possibilities, we pick the one with minimal $y^*$ and then with minimal $x^*$. We then define $M[i,j] = \prev{i-1}{M[1,j]}$. In other words, below every pair in $M$ we put its predecessor. Observe that the properties of predecessors immediately imply $\lcisto{M[i,j]} = \lcisto{M[1,j]}-i+1 = \ell - i - j + 2$.

If we pick, instead of some $b_j$, another $b'_j$ such that $\lcis{a,b'_j} = \lcis{a,b_j} = \ell - j + 1$, we will get exactly the same $M[1,j]$, and thus the same $M[i,j]$ for all $i = 1, 2, \ldots, n$ -- this is because of the tie-breaker rule for the choice of $M[1,j]$. Thus, the matrix $M$ does not depend on the choice of $b_1, \ldots, b_n$, but only on $\hat{A}$ and $\hat{B}$.

We begin with a technical lemma about $M$ which will be useful later. This observation is a generalization of the idea introduced in Section \ref{section:idea} -- if $s$ different significant pairs are incident to a single vertex $x \in \hat{A}$, then among their $i$-th predecessors we expect at least about $s/i$ distinct values.

\begin{lemma}
\label{lemma:matrix-distinct}
 For some $i, i', j, j'$ with $1 \leq j < j' \leq n$ and $1 \leq i, i' \leq n$, let $(x,y) = M[i,j]$ and $(x',y') = M[i',j']$. If $j' \geq j+i$, then $y' < y$.
 \begin{proof}
  Suppose to the contrary that $y \leq y'$. As $y' \leq b_{j'}$ and $a$ is the last element of $\hat{A}$, it would imply $(x,y) \leq (a,b_{j'})$, which would in turn yield $\lcis{a,{b_{j'}}} \geq \lcisto{x,y} = \lcisto{M[i,j]} = \lcisto{M[1,j]} - i + 1 = \lcis{a,b_j} - i + 1$. But as $j' \geq j+i$, there must be $\lcis{a,b_{j'}} \leq \lcis{a,b_j} - i$. This contradiction shows $y' < y$.
 \end{proof}

\end{lemma}

As stated before, we will refer to the symbols of $\hat{A}$ and $\hat{B}$ as \emph{colors}, imagining that every entry $(x,y)$ in $M$ is painted with a color corresponding to $\col{x,y}$, the (common) symbol of $\hat{A}[x]$ and $\hat{B}[y]$. In every column of $M$ the colors are strictly decreasing, and thus different. Hence, no color can have more than $n$ entries in $M$. It is also evident that two entries in $M$ must correspond to different elements in $\hat{A}$ and $\hat{B}$ if they have different colors. The main lemma of this section states, roughly, that the rows of $M$ do not contain too few colors:

\begin{lemma}
\label{lemma:submatrix}
 There is some $2 \leq k \leq n/\delta$ such that every submatrix of $M$ consisting of some $\lceil\frac{n}{\delta}\rceil$ columns (not necessarily consecutive) and rows $1, \ldots, k-1$ uses at least $\lceil k \delta \rceil$ colors.
 \begin{proof}
 Consider all $k$ that are powers of $2$: $k = 2^q$ for $1 \leq q \leq \lfloor \log{n} - \log{\delta} \rfloor$. Suppose, to the contrary, that for every such $k = 2^q$ we can find some $\lceil\frac{n}{\delta}\rceil$ columns $c_1, \ldots, c_{\lceil\frac{n}{\delta}\rceil}$ of $M$ which have at most $\delta \cdot 2^q$ colors in total in rows $1, 2, \ldots, 2^q -1$. From these columns $c_i$ and the lower half of these rows ($2^{q-1},\ldots,2^q-1$) we form a submatrix $M_q$ of $M$. These matrices are defined for $1 \leq q \leq \lfloor \log{n} - \log{\delta} \rfloor$ and have the following properties: 
 \begin{itemize}
     \item they are disjoint submatrices of $M$ (as every one takes different rows),
     \item for any $q$, the matrix $M_q$ contains $2^{q-1} \cdot \lceil \frac{n}{\delta} \rceil$ pairs,
     \item for any $q$, the entries of $M_q$ use at most $\delta \cdot 2^q$ colors between them.
 \end{itemize}

 Let a color be \emph{$q$-strong}, if at least $\frac{n}{4 \delta^2}$ entries in $M_q$ are of that color. Observe that a particular color can be $q$-strong for at most $4 \delta^2$ distinct values of $q$, otherwise -- as all $M_q$'s are disjoint -- there would be more than $n$ entries of that color in $M$, which is impossible.

For any $q$, the colors which are not $q$-strong can make up for at most half of entries in $M_q$ (as there are $\delta \cdot 2^q$ colors in $M_q$, none of which can have more than $\frac{n}{4 \delta^2}$ entries -- a total of $\frac{n}{2\delta} \cdot 2^{q-1}$). Hence, there are at least $\frac{n 2^{q-1}}{2 \delta}$ pairs in $M_q$ which have $q$-strong colors. Let us \emph{mark} all these entries of $M_q$.

For any $t = 0, 1, \ldots, 2^q-1$ let $\mathcal{M}_q(t)$ be the set of columns $M_q[\cdot, j]$ with $j \equiv t \mod 2^q$. For a fixed $q$, at least one of the sets $\mathcal{M}_q(t)$ must contain at least $\frac{n}{4 \delta}$ marked entries, as there are $\frac{n 2^{q-1}}{2 \delta}$ marked entries in $M_q$ split between $2^{q}$ sets. But we can show that if $(x,y)$ and $(x',y')$ are two different pairs in some $\mathcal{M}_q(t)$, then $y \neq y'$. Indeed, both pairs are either in the same column -- which makes them have different colors and thus no common elements -- or at least $2^q$ columns apart, in which case $y \neq y'$ because of Lemma \ref{lemma:matrix-distinct}. This in turn means that for every $q$ there is a set $B_q$ of at least $\frac{n}{4 \delta}$ distinct elements of $\hat{B}$, each of a $q$-strong color. A color can be $q$-strong for at most $4\delta^2$ values of $q$, so in the sum $B_1 \cup \ldots \cup B_{\lfloor \log{n} - \log \delta \rfloor}$ every element can be repeated at most $4 \delta^2$ times. This accounts for at least $\lfloor\log n - \log \delta \rfloor \cdot \frac{n}{4 \delta} \cdot \frac{1}{4 \delta^2} = n \frac{\lfloor \log n - \log \delta  \rfloor}{16 \delta^3}$ distinct elements of $\hat{B}$. For $\delta = \frac{\log^{1/3}{n}}{4}$ this is equal to $4n \cdot \frac{\lfloor \log n - 1/3 \cdot \log \log n + 2 \rfloor }{\log n} > 3n \geq |\hat{B}|$. The contradiction proves that at least one $k = 2^q$ for some $q$ must satisfy the statement.
\end{proof}
\end{lemma}

Now let us return to the graph $G_\sigma$ of significant pairs. Recall that $\sigma$ is the largest symbol appearing in input sequences -- for the rest of the section, we retain this assumption. As Figure \ref{figure:example-significant} shows, any two incident edges must correspond to pairs with different values of LCIS, as otherwise the pairs could not be significant. This is formalized in a following simple observation:

\begin{lemma}
\label{lemma:sig-between}
\begin{enumerate}[a)]
 \item If $(x,y_1), (x,y_2), \ldots, (x,y_s)$ are significant $\sigma$-pairs for $y_1 > y_2 > \ldots > y_s$, then for every $1\leq i \leq j \leq s$, we have $\lcis{x,y_i} \geq \lcis{x,y_j} + (j-i)$. 
 \item If $(x_1,y), (x_2,y), \ldots, (x_s,y)$ are significant $\sigma$-pairs for $x_1 > x_2 > \ldots > x_s$, then for every $1\leq i \leq j \leq s$, we have $\lcis{x_i,y} \geq \lcis{x_j,y} + (j-i)$. 
\end{enumerate}
 \begin{proof}
  The first claim easily follows from the fact that $\lcis{x,y_i} \geq \lcis{x,y_{i+1}} + 1$, which is part of the definition of the significant pair. The second proof is symmetric.
 \end{proof}

\end{lemma}

Finally, we can restate Lemma \ref{lemma:no-bad} and prove it using predecessor matrices:

\LemmaNoBad*

\begin{proof}
  Assume, to the contrary, that for every $k$ there is a $k$-dense suffix. Pick an arbitrary $k \leq n/\delta$ and let $x \in A$ be such that $A_x^\to$ is $k$-dense. Let $|\hat{A}| = a$, let $r = \lceil n/\delta \rceil$ and let $(x, c_1), \ldots, (x, c_r)$ be the significant pairs from the definition of $k$-dense suffix (meaning that each $c_i$ has $k$ neighbours in $A_x^\to$). We can assume that $c_1 > c_2 > \ldots > c_r$. We will now show another sequence $c'_1, c'_2, \ldots, c'_r$ such that:
  \begin{enumerate}[(1)]
  \item $\lcis{a,c'_1} > \lcis{a,c'_2} > \ldots > \lcis{a,c'_r},$
  \item $\lcis{a,c'_i} \geq \lcis{x,c'_i} + k-1\mbox{ for } i = 1, 2, \ldots, r.$
  \end{enumerate}
  
  To do that, we set $c'_1 = c_1$, and for $i > 1$, we pick $c'_{i+1} = c_{i+1}$ if $\lcis{a,c_{i+1}} < \lcis{a,c'_i}$. If not, we define $c'_{i+1}$ to be the largest element with $\lcis{a,c'_{i+1}} = \lcis{a,c'_{i}} - 1$. Observe that in the second case we know that $\lcis{a,c_{i+1}} \geq \lcis{a,c'_i} > \lcis{a,c'_{i+1}}$, so always $c'_{i+1} \leq c_{i+1}$.
  
  Inequality (1) follows immediately from the definition of $c'_i$. To see (2), first observe that if $c'_{i} = c_{i}$, then there are $k$ significant $\sigma$-pairs between $(x,c_i)$ and $(a,c_i)$ -- neighbors of $c_i$ -- which we will denote by $(y_1,c_i),\ldots, (y_k,c_i)$ and assume that $y_1 > \ldots > y_k$. As we assume $\sigma$ to be the largest symbol, we can write $\lcis{y_j,c_i} = \lcisto{y_j,c_i}$. From this and Lemma \ref{lemma:sig-between} we derive:
  
    $$\lcis{a,c_i} \geq \lcis{y_1,c_i} \geq \lcis{y_k,c_i} + k-1 \geq \lcis{x,c_i} + k-1.$$
    
   Now consider the case $c'_i < c_i$. Let $\beta$ be the smallest integer such that $\beta < i$ and $c_{i-\beta} = c'_{i-\beta}$ (it always exists, as we can take $\beta = i-1$). From the definition of $c'_i$ we have $\lcis{a,c'_i} = \lcis{a,c'_{i-1}}-1 = \ldots = \lcis{a,c'_{i-\beta}}-\beta = \lcis{a,c_{i-\beta}}-\beta \geq \lcis{x,c_{i-\beta}} + k-1 - \beta$. Now, because all $(x,c_j)$ are significant, we have $\lcis{x,c_{i-\beta}} \geq \lcis{x,c_{i}} + \beta$ from Lemma \ref{lemma:sig-between}, so $\lcis{a,c'_i} \geq \lcis{x,c_{i}} + k-1 \geq \lcis{x,c'_{i}} + k-1$.

  The pairs $(a,c'_i)$ are some choice of $r$ different columns of the predecessor matrix $M$. Consider any $c'_i$, and its corresponding column $j$. Pick a positive integer $s \leq k-1$. Let $(a^*_i,c^*_i) = M[s,j]$. Recall that from the definition of $M$ we have $\lcisto{a^*_i,c^*_i} = \lcis{a,c'_i} - s + 1$. We also know that $c^*_i \leq c'_i \leq c_i$. If $a^*_i < x$, then $(a^*_i,c^*_i) \leq (x,c'_i)$, which implies $\lcis{a,c'_i} - s + 1 = \lcis{a^*_i,c^*_i} \leq \lcis{x,c'_i} \leq \lcis{a,c'_i} - k + 1$, which is impossible for $s \leq k-1$. Then $a^*_i \geq x$. So the only colors available for $M[s,j]$ for the chosen $r$ columns and $s \leq k-1$ are those appearing in $A_x^\to$, and there are at most $\lceil k \delta \rceil$ of them. Hence, for any $k$ we can produce, from a $k$-dense suffix, an $\lceil \frac{n}{\delta} \rceil$-column submatrix of $M$ having at most $\lceil k \delta \rceil$ colors in total in its first $k-1$ rows. This contradicts Lemma \ref{lemma:submatrix} and proves that for some $k$ there are no $k$-dense suffixes.
\end{proof}

\section{The algorithm}
\label{section:algo-full}

To implement our algorithm, we need a specific data structure -- an associative array which can store a number of elements ordered by their \emph{keys}. We assume the keys to be distinct integers between $1$ and $n$. This data structure $A$ must provide the following operations:
\begin{itemize}
 \item $\fun{Insert}{A,s}$ -- adds the element $s$ to $A$,
 \item $\fun{Delete}{A,x}$ -- removes the element having key $x$ from the $A$ (we assume that this is called only for $x \in A$ ),
 \item $\fun{Find}{A,x}$ -- returns the element whose key is $x$ if there is one, or NULL otherwise,
 \item $\fun{Next}{A,x}$, $\fun{Prev}{A,x}$ -- returns the first element whose key is larger (respectively, smaller) than $x$.
\end{itemize}

To achieve the desired running time, we need all these operations to work in $\O{\log \log n}$ complexity (possibly amortized), and \emph{van Emde Boas queue} \cite{vanEmdeBoas75} does exactly that. While the standard implementation requires $\O{n}$ space (and thus $\O{n}$ initialization time, which would be too much for us, as we employ $\O{n}$ queues), there is also a randomized version (\cite{MehlhornN90}) that needs only $\O{m}$ time and space, where $m$ is the maximal number of elements on the queue. In \shortversion{the full version of the paper}\fullversion{Appendix \ref{sec:vEB}} we show how to construct a deterministic van Emde Boas queue with $\O{m + \frac{n}{\log^{c} n}}$ time and space bounds, with $c$ being any desired constant, while retaining the $\O{\log \log n}$ query complexity. These bounds also suit our needs.

The algorithm takes, as the input, two integer sequences $A$ and $B$ with $|A| = |B| = n$. Its main idea is to consider all symbols from $A$ and $B$ in increasing order (there are at most $2n$ of them, so we can sort them in $\O{n \log n}$). For every symbol $\sigma$, the algorithm finds and stores all significant $\sigma$-pairs. For that, we employ $n$ van Emde Boas queues $Q_1, Q_2, \ldots, Q_n$, with every $Q_k$ storing the significant pairs $(x,y)$ with $\lcisto{x,y} = k$, sorted by $x$. For convenience, we define $Q = Q_1 \cup \ldots \cup Q_n$. We also keep $Q_0$ as one-element queue $(0,0)$. 

Whenever some $Q_k$ contains two pairs $(x,y) \leq (x',y')$, we drop $(x',y')$ from $Q_k$. Informally, we can do it because $(x,y)$ can replace $(x',y')$ in every situation. We say that $(x,y)$ \emph{dominates} $(x',y')$ and remove any dominated pairs from any $Q_k$. Observe that a pair is significant if and only if it is not dominated by any pair of the same symbol (a significant pair, however, may still be dominated by other pairs with larger symbols).

This leads to the following invariant: 

\begin{claim}[Algorithm invariant]
\label{thm:alg-inv}
For every $k$, all pairs $(x,y) \in Q_k$ are in strict increasing order with respect to $x$ and in strict decreasing order with respect to $y$. 
\end{claim}

To keep the invariant, we modify $\fun{Insert}{}$ into the following $\fun{Insert-Inv}{}$ procedure. It only inserts a pair $(x,y)$ if it is not dominated by another pair, and after inserting it removes all larger pairs.

\begin{algorithm}
\caption{New version of $\fun{Insert}{}$ keeping the invariant}
 \begin{algorithmic}[1]
    \Procedure{Insert-Inv}{$Q_k, (x,y)$}
        \State $(x',y') \gets$ \Call{Find}{$Q_k, x$} \Comment{check for other pairs with key $x$}
        \If {$(x',y') \neq$ \textbf{null} and $y' \leq y$}
            \State \textbf{return}
        \EndIf
        \If {$(x',y') \neq$ \textbf{null} and $y' > y$}
            \State \Call{Delete}{$x'$}
        \EndIf
        \State $(a,b) \gets$ \Call{Prev}{$Q_k, x$}
        \If {$(a,b) \neq$ \textbf{null} and $b \leq x$} \Comment{$(a,b) \leq (x,y)$, so we should not insert $(x,y)$}
            \State \textbf{return}
        \EndIf
        \State \Call{Insert}{$Q_k,(x,y)$}
        \Repeat \label{alg:insinv:loopstart} \Comment{now we remove all $(a,b) \geq (x,y)$, restoring the invariant} 
            \State $(a,b) \gets$ \Call{Next}{$Q_k, x$}
            \If {$(a,b) \neq$ \textbf{null} and $\geq (x,y)$}
                \State \Call{Delete}{$Q_k,a$}
            \EndIf
        \Until{$(a,b)$ = \textbf{null} \textbf{or} \textbf{not} $(a,b) \geq (x,y)$} \label{alg:insinv:loopend}
    \EndProcedure
  \end{algorithmic}  
\end{algorithm}
  
The amortized complexity of $\fun{Insert-Inv}{}$ is $\O{\log \log n}$: the loop in lines \ref{alg:insinv:loopstart}-\ref{alg:insinv:loopend} deletes an element with every iteration (so it cannot do more iterations than the total number of elements in queue), and outside the loop there is only a constant number of standard queue operations.
  
Apart from queues $Q_1, Q_2, \ldots, Q_n$ we will also need, for every symbol $\sigma$, two van Emde Boas queues $X_\sigma$ and $Y_\sigma$ which store positions of all $\sigma$-symbols in $A$ and $B$, respectively: $X_\sigma = \{i : A[i] = \sigma\}$, $Y_\sigma = \{j : B[j] = \sigma\}$. These structures do not change during the algorithm, and their sole purpose is finding $\sigma$-symbols closest to a given position. 

Now we are ready to introduce the main idea of the algorithm. Recall that we assume that the number of significant pairs between $A$ and $B$ is at most $\O{\frac{n^2}{t}}$  with $t = t(n) = \Th{\log^p n}$ for some $p$. We iterate over all the symbols, dividing them into two categories:
\begin{itemize}
 \item \emph{frequent} -- appearing more than $\frac{n}{\sqrt{t}}$ times in $B$,
 \item \emph{infrequent} -- with at most $\frac{n}{\sqrt{t}}$ occurrences in $B$.\footnote[3]{The technique of splitting symbols of a string according to their number of occurences is not new -- it has been used, e.g. in \cite{Abrahamson87} and \cite{AmirLP00}, though it is more common to have split thresholds closer to $\sqrt{n}$.}
\end{itemize}

Let us start with an informal sketch of the algorithm behavior for both cases. The frequent symbols are easier: for every such symbol $\sigma$ we iterate through all previously found pairs, and for every $(x,y) \in Q$ we find the next occurrence of $\sigma$ after $A[x]$ (say, $A[x^*]$) and the next occurrence $y^*$ of $\sigma$ in $B$ after $y$. In other words, we find a $\sigma$-pair $(x^*, y^*)$ for which $(x,y)$ is a predecessor. As we will ensure that $Q$ contains only significant pairs (and thus cannot get too big) and there are no more than $\sqrt{t}$ frequent symbols, the total complexity will fit into desired limits.

To handle infrequent symbols, observe that every such symbol in $A$ can form a matching pair with at most $\frac{n}{\sqrt{t}}$ elements of $B$. Hence, there are at most $\frac{n^2}{\sqrt{t}}$ matching pairs on infrequent symbols, so we can iterate through all of them. The hardest part is to determine, for every infrequent pair $(x,y)$, the value of $\lcisto{x,y}$. For that, we will need a separate subroutine and a non-trivial analysis.

The whole algorithm is presented below:

\begin{algorithm}[H]
\caption{LCIS by significant pairs}
\begin{algorithmic}[1]
\Procedure{LCIS}{$A,B$}
\State $Q_0 \gets \{(0,0)\}$
\ForAll {$\sigma$ -- symbols in increasing order}
    \State $T \gets \varnothing$ \Comment{for storing new pairs}
    \If {$\sigma$ occurs less than $n/\sqrt{t}$ times in $B$} \Comment{if $\sigma$ is infrequent\ldots}
        \ForAll { $x : A[x] = \sigma$, in increasing order } \Comment{fix $x$\ldots}
            \State $k \gets 0$ 
            \ForAll {$y : B[y] = \sigma$, in inc. order } \Comment{\ldots compute $\lcisto{x,y}$ for all $y$}
                \State $k' \gets$ \Call{ComputeNextPair}{$x,y,k$} \Comment{using a special subroutine}
                \State $T \gets T \cup (x,y,k')$
                \State $k \gets k'$ \Comment{$k = \lcisto{x,y}$, for last considered $y$}
            \EndFor
        \EndFor
    \Else	\Comment{If $\sigma$ is frequent\ldots}
            \For{$k = 1, 2, \ldots, n$}
                \ForAll {$(x,y) \in Q_k$}	\Comment{every pair $(x,y) \in Q$ may be a predecessor\ldots}
                    \State $x' \gets$ \Call{Next}{$X_\sigma, x$}
                    \State $y' \gets$ \Call{Next}{$Y_\sigma, y$} \Comment{\ldots of some $\sigma$-pair $(x',y')$}
                    \State $T \gets T \cup (x',y',k+1)$
                \EndFor
            \EndFor
    \EndIf
    \ForAll {$(x,y,k) \in T$}		\Comment{all new pairs are now added to $Q$}
            \State \Call{Insert-Inv}{$Q_k, (x,y)$}
    \EndFor
\EndFor
\State\textbf{return} largest $k$ with $Q_k \neq \varnothing$
\EndProcedure
\end{algorithmic}
\end{algorithm}

Before analyzing the algorithm, we must explain the \fun{ComputeNextPair}{} subroutine. 
It takes three arguments: positions $x \in A$, $y \in B$, such that $A[x] = B[y] = \sigma$, and an integer $k$. It assumes that $\lcisto{x,y} \geq k$ and its goal is to find the exact value of $\lcisto{x,y}$. It also assumes that for every $j < \lcisto{x,y}$, there is a pair $(x_j,y_j) \in Q_j$ with $(x_j, y_j) \prec (x,y)$ -- informally, this means that all the predecessors of $(x,y)$ have already been considered, and Lemma \ref{lemma:correctness} will prove that this condition is indeed satisfied whenever \fun{ComputeNextPair}{} is invoked.

Therefore the subroutine must determine the largest $\ell \geq k-1$ for which there is a pair $(x',y') \prec (x,y)$ with $\lcisto{x',y'} = \ell$. We can guess $\ell$, and verify whether there exists a right pair $(x',y') \prec (x,y)$ in $Q_{\ell}$: if there is one, then $\fun{Prev}{Q_{\ell},x} \leq (x,y)$. This allows us to do a binary search for $\ell$:

\begin{algorithm}[H]
\caption{Finding $\lcisto{x,y}$, with assumption that it is at least $k$}
 \begin{algorithmic}[1]
    \Procedure{ComputeNextPair}{$x,y,k$}
    \State $d \gets 1$	\Comment{first, find $d$ -- a rough approximation for $\ell - k$}
    \While{\Call{Prev}{$Q_{k+d},x$} $ \prec (x,y)$}
        \State $d \gets 2d$	\Comment{if $d$ is too small, try $2d$}
    \EndWhile
    \State $p \gets k$		\Comment{now we know that $k + d/2 \leq \ell < k + d$}
    \State $q \gets k+d$
    \While{$p<q$}		\Comment{so we can do the real binary search}
        \State $s \gets \lceil \frac{p+q}{2} \rceil$
        \If{\Call{Prev}{$Q_{s},x$} $ \prec (x,y)$}        
            \State $p \gets s$
        \Else
            \State $q \gets s-1$
        \EndIf
    \EndWhile
    \State \textbf{return} $p$
    \EndProcedure
  \end{algorithmic}
\end{algorithm}

The first part of the algorithm finds $d$ for which $d/2 \leq |\ell - k| < d$ (if $\ell = k$, then we assume $d = 1$). The second one is a binary search on the interval $[k, k+d)$. Therefore, the algorithm makes at most $2\log(d+1) = 2 \cdot \log (\lcisto{x,y} - k + 2) $ steps, each step invoking a queue operation once.

Let us now go back to the main algorithm, proving its correctness: 

\begin{lemma}\label{lemma:correctness}
 After processing a symbol $\sigma$, the following two facts hold:
 \begin{enumerate}[(1)]
  \item If $(x,y) \in Q_k$, then $(x,y)$ is a significant $\sigma'$-pair for some $\sigma' \leq \sigma$ with $\lcisto{x,y} = k$;
  \item For any $k \geq 1$, $\sigma' \leq \sigma$ and for every $\sigma'$-pair $(x,y)$ with $\lcisto{x,y} \geq k$, there is some $(x^*,y^*) \in Q_k$ with $(x^*,y^*) \leq (x,y)$.
 \end{enumerate}

 \begin{proof}
  Let us use induction on $\sigma$. Observe that assuming induction hypothesis, we only need to prove two weaker facts:
  
  \begin{enumerate}[(1')]
  \item If $(x,y) \in Q_k$, then $\lcisto{x,y} = k$;
  \item For any $k \geq 1$ and for every $\sigma$-pair $(x,y)$ with $\lcisto{x,y} = k$, there is some $(x^*,y^*) \in Q_k$ with $(x^*,y^*) \leq (x,y)$.
 \end{enumerate}

 Indeed, for any pair $(x,y)$ with $\lcisto{x,y} = k' > k$, (2) is true because of the induction hypothesis applied to the pair $(x',y') = \prev{k'-k}{x,y}$. We know that $(x',y')$ is a $\tau$-pair for some $\tau < \sigma$, so induction hypothesis provides a pair $(x^*,y^*) \in Q_k$ with $(x^*, y^*) \leq (x',y') \leq (x,y)$. Next, note that (2) is also true for $\sigma' < \sigma$, again from induction hypothesis and the fact that once a pair is in $Q$, it can only be dislocated by another pair that dominates it.
  
  Also, if we show (2) and (1'), this will automatically imply that every $(x,y) \in Q_k$ must be significant, and thus that (1) is true -- let $(x',y')$ be a pair with $(x',y') \leq (x,y)$ and $\lcisto{x',y'} = \lcisto{x,y} = k$. There is, by (2), another pair $(x'',y'') \leq (x',y') \leq (x,y)$, which is also in $Q_k$. But with Claim \ref{thm:alg-inv}, $(x,y)$ and $(x'',y'')$ cannot both be in $Q_k$ unless $(x,y) = (x',y') = (x'',y'')$, so $(x,y)$ is significant.
  
  To prove the remaining statements (1') and (2'), consider two cases:
  
  \textbf{Case $1$: infrequent $\sigma$}. Pick an arbitrary $\sigma$-pair $(x,y)$ and let $\lcisto{x,y} = k$. We will show that at some point during processing symbol $\sigma$ the instruction \fun{Insert-Inv}{$Q_k,(x,y)$} is invoked.
  
  For any integer $i$ with $1 \leq i \leq k$ let $(x_i,y_i) = \prev{i}{x,y}$. We know that $\col{x_i,y_i} < \sigma$ and $\lcisto{x_i,y_i} = k-i$, so by induction hypothesis (2) there was some $(x_i^*,y_i^*) \in Q_{k-i}$ with $(x_i^*,y_i^*) \leq (x_i,y_i)$, which implies $(x_i^*,y_i^*) \prec (x,y)$ (observe that for $i = k$, we need the artificial pair $(0,0) \in Q_0$). But then when $(x,y)$ is considered by \fun{ComputeNextPair}{}, for every $i \geq 1$ and for each of the predecessors $(x_i,y_i) = \prev{i}{x,y}$ there is a pair $(x_i^*, y_i^*) \leq (x_i, y_i)$ in $Q_{k-i}$. This satisfies the conditions needed by \fun{ComputeNextPair}{}, so the binary search properly computes $\lcisto{x,y}$ as $k$ (note that it is not possible to find any larger candidate for predecessor -- any $(u,v) \prec (x,y)$ found in $Q_{k'}$ for $k' \geq k$ would mean either that $\lcisto{u,v}$ had been computed wrong, or that $\lcisto{x,y} > k$). This shows that the algorithm tries to insert $(x,y)$ into the proper queue $Q_k$, which immediately proves (1'). Also, $(x,y)$ may remain in $Q_k$, be dislocated later, or even fail to be inserted because of some other pair dominating it. Either way, some pair $(x^*,y^*) \leq (x,y)$ will be present in $Q_k$ to the very end of the algorithm. This completes the proof of (2').

  \textbf{Case $2$: frequent $\sigma$}. Let us first prove (2') for any $k$ and for any $\sigma$-pair $(x,y)$ with $\lcisto{x,y} = k$. We can assume that $(x,y)$ is significant (if not, we replace it with a significant $\sigma$-pair $(x',y') \leq (x,y)$). As in Case 1, from the induction hypothesis we know that some pair $(x',y') \prec (x,y)$ is present in $Q_{k-1}$. The algorithm must at some point consider $(x',y')$. If the next $\sigma$ symbol in $A$ after $x'$ is not $x$ but some $z$, then $\lcisto{z,y} \geq k$, so $(x,y)$ could not be significant. By the same argument, the next $\sigma$ symbol in $B$ after $y'$ must be $y$. Therefore $(x,y)$ is a candidate to be inserted into $Q_k$, so either it remains there itself, or is dominated by other pair $(x^*,y^*) \in Q_k$. Either way, (2') is shown.
  
  For (1) we need to rule out a possibility that a $\sigma$-pair $(x,y)$ with $\lcis{x,y} = k$ will be inserted, besides $Q_k$, into some other $Q_{k'}$ with $k' \neq k$, as a frequent pair could be theoretically considered multiple times by the algorithm. But for $k' > k$ this would have been caused by another pair $(x^*,y^*) \in Q_{k'-1}$ with  $(x^*,y^*) \prec (x,y)$. This contradicts $\lcisto{x,y} = k$, as $k'-1 \geq k$. For $k'<k$ observe that by induction hypothesis (2) applied to $\prev{k-k'}{x,y}$ we already have a pair in $Q_{k'}$ which dominates $(x,y)$, so the insertion must fail.
\end{proof}
\end{lemma}

\begin{corollary}
 The algorithm correctly returns the length of longest common increasing subsequence of $A$ and $B$.
 \begin{proof}
  Let $k$ be the value of LCIS and let $(x,y)$ be a significant pair with $\lcisto{x,y} = k$. From statement (2) of Lemma \ref{lemma:correctness} we know that some pair $(x',y') \leq (x,y)$ must be in $Q_k$ at the end of the algorithm. Therefore the algorithm returns at least $k$. On the other hand, for every $k' > k$,  $Q_{k'} = \varnothing$, as any pair in it would contradict statement (1) from Lemma \ref{lemma:correctness}.
 \end{proof}
\end{corollary}

\begin{lemma} \label{lemma:next-pair-complexity}
 Let $x \leq |A|$ and $(x,y_1)$, $(x,y_2)$, \ldots, $(x,y_m)$ be all $\sigma$-pairs formed by $x$, for an infrequent $\sigma$. Then, all calls to $\fun{ComputeNextPair}{x,\cdot,\cdot}$ work in $\O{\frac{n (\log \log n)^2} {\sqrt{t}} }$ total time complexity.
 \begin{proof}
  We know that $m \leq \frac{n}{\sqrt{t}}$, as $\sigma$ is infrequent. Recall also that $\log t =  \Th{\log \log n}$. Let $\ell_i = \lcisto{x,y_{i}} - \lcisto{x,y_{i-1}} + 2$ for $i = 1, 2, \ldots, m$, assuming $y_0 = 0$. The $i$-th call to $\fun{ComputeNextPair}$ requires $\O{\log \ell_i}$ steps of binary search, with every step having $\O{\log \log n}$ complexity from queue operations. Therefore, the whole procedure works in $\O{\log \log n \cdot \sum_{i=1}^{m} \log{\ell_i}}$. Consider two cases:
  \begin{itemize}
   \item If $m \leq \frac{n}{\sqrt{t} \log n}$, then $\sum_{i=1}^{m} \log{\ell_i} \leq m \log n = \O{n/\sqrt{t}}$.
   \item If $m > \frac{n}{\sqrt{t} \log n}$, then the Jensen equality yields:
   
   $\sum_{i=1}^{m} \log{\ell_i} = m \frac{\sum_{i=1}^{m} \log{\ell_i}}{m} \leq m \log \left( \frac{\sum_{i=1}^{m} \ell_i}{m} \right)$,
   
   and as $\sum \ell_i = 2m + \lcisto{x,y_{m}} - \lcisto{x,y_0} \leq 3n$ we have:
   
   $\sum_{i=1}^{m} \log{\ell_i} \leq m \log \frac{3n}{m} < m \log (3 \sqrt{t} \cdot \log n) = \O{m \log \log n}.$
  \end{itemize}
  With $m \leq n/\sqrt{t}$, the total complexity in both cases is $\O{\frac{n (\log \log n)^2} {\sqrt{t}} }$.
 \end{proof}
\end{lemma}

\begin{proof}[Proof of Theorem \ref{thm:main-algo}]
We already know that the algorithm correctly computes $\lcis{A,B}$. It remains to determine its complexity.

There are $\O{n}$ van Emde Boas queues. Each of them is initialized in constant time if we use randomized version \cite{MehlhornN90}, or in $\O{n / \log^c{n}}$ with some $c$ large enough if we choose the version \shortversion{described in the appendix of the full version of the paper}\fullversion{from Appendix \ref{sec:vEB}}. Either way, total complexity is $\O{n^2 / \log^c{n}}$, which is fast enough.

We first analyze the cost for infrequent symbols -- it is dominated by the calls of $\fun{ComputeNextPair}{}$. By Lemma \ref{lemma:next-pair-complexity} the cost is $\O{\frac{n (\log \log n)^2} {\sqrt{t}} }$ for a fixed $x$, which yields $\O{\frac{n^2 (\log \log n)^2} {\sqrt{t}}}$ total complexity.

Now let us move on to frequent symbols. For every such symbol, we iterate over all $Q$. But from Lemma \ref{lemma:correctness} we know $Q$ only contains significant pairs, therefore $|Q| = \O{\frac{n^2}{t}}$. As every iteration needs only a constant number of queue operations, the total cost for a single symbol is $\O{\frac{n^2 \log \log n}{t} }$.

Finally, observe that there are at most $\sqrt{t}$ frequent symbols (otherwise there would be $|B| > n$), so the final complexity in this case is $\O{\frac{n^2 \log \log n}{\sqrt{t}}}$.

It is also worth noting that we can replace the threshold between frequent and infrequent symbols $\frac{n}{\sqrt{t}}$ with $\frac{n}{\sqrt{t \log \log n}}$, and all the proofs would essentially work in the same way, with only minor changes needed. This way we could show the complexity of the algorithm to be in fact $\O{\frac{n^2 (\log \log n)^{3/2}}{\sqrt{t}}}$. The current analysis seems, however, a bit easier to read.

\end{proof}

\section{Final remarks and open problems related to LCIS}

We have shown an algorithm for LCIS that breaks the $\O{n^2}$ barrier, but there still appears to be plenty of room for improvement and further work on this matter. First, the bound in Theorem \ref{thm:main-count} for the number of significant pairs may not be tight. In \shortversion{the full version of the paper}\fullversion{Appendix \ref{sec:howmanypairs}} we give an example of two sequences $A$ and $B$ having $\Om{\frac{n^2}{\log n}}$ significant pairs, but this still leaves a gap between $\Om{\frac{n^2}{\log n}}$ and $\O{\frac{n^2}{(\log n)^{1/3}}}$. Also, the algorithm itself might be improved to work in $\O{s \cdot (\log \log n)^k }$, where $s$ is the number of significant pairs and $k$ is a constant. Taking all this into account, we conjecture that there is an $\O{\frac{n^2 (\log \log n)^k}{\log n}}$ algorithm which uses the significant pairs technique.

The second question related to LCIS stems from the papers \cite{AbboudHWW16} and \cite{AbboudB18}: we know that there is a constant $c \leq 7$ such that an $\O{\frac{n^2}{\log^c{n}}}$ algorithm for LCS would lead to unexpected breakthroughs in circuit complexity. Can a similar statement be made for LCIS?

\bibliography{lcis-log}

\clearpage 
\appendix
\section{A modified van Emde Boas queue}\label{sec:vEB}

In this section we show how to modify van Emde Boas queue -- for some given constant $c$, we want it to use $\O{m + \frac{n}{\log^c{n}}}$ space, where $n$ is the universe size (the elements' keys are integers between $1$ and $n$), and $m$ is the number of queue elements. Our version retains $\O{\log \log n}$ worst-case complexity of all operations.

For convenience, assume that $n = 2^{2r}$ for some integer $r$ and let $s = 2^r = \sqrt{n}$. For any integer $x$ with $x \leq n$ let:
\begin{itemize}
 \item $\lo{x} = (x \mod s)$ be the number consisting of $m$ least significant bits of $x$,
 \item $\hi{x} = \lfloor x/s \rfloor$ the number consisting of $m$ most significant bits of $x$. 
\end{itemize}

Thus $x = \lo{x} + \hi{x} \cdot s$. Recall that a standard van Emde Boas queue $T$ of size $n$ consists of:

\begin{itemize}
 \item recursive queues $U_1, \ldots, U_s$ of size $s$ -- each $U_i$ stores the elements with keys $x$ for which $\hi{x} = i$;
 \item a single structure $H$ which stores the indices $i$ for which $U_i$'s are nonempty;
 \item a single queue $V$ of size $s$, storing the indices of nonempty $U_i$'s -- the same data as in $H$, but stored differently, in a recursive queue;
 \item two integers $\fun{Min}{T}$ and $\fun{Max}{T}$, keeping the minimal and the maximal element of $T$. Those elements are not part of any $U_i$.
\end{itemize}

We will analyze the $\fun{Next}{T,x}$ operation for $x \notin T$ -- the other ones are very similar. This operation works as follows:
\begin{itemize}
 \item Let $i = \hi{x}$ and check in $H$ if $U_i$ is nonempty;
 \item If $U_i$ is empty, we find the next nonempty $U_{i'}$ for some $i' > i$, by a single call $i' = \fun{Next}{V,i}$. Then the answer is $\fun{Min}{U_{i'}}$, which we can find in constant time;
 \item If $U_i$ is nonempty but $x = \fun{Max}{U_i}$, we do as before -- find next nonempty $U_{i'}$ for $i' > i$ and return $\fun{Min}{U_{i'}}$;
 \item If $U_i$ is nonempty and $x < \fun{Max}{U_i}$, the answer is $\fun{Next}{U_i,\lo{x}}$.
\end{itemize}

In any case we do one operation on $H$, a single recursive call (either to $V$ or some $U_i$) and some constant-time calls to $\fun{Min}$ or $\fun{Max}$. The original implementation of van Emde Boas \cite{vanEmdeBoas75} uses a simple $0/1$ array as $H$ (i.e. $H[i] = 1$ iff $U_i$ is nonempty). This is the easiest approach, but $H$ needs $\sqrt{n}$ bits on the first recursion level, $n^{3/4}$ on the second, $n^{7/8}$ on the third, and so on, up to $n$ bits at the deepest level of recursion. The total space needed can go up to $\O{n}$, and all $H$'s need initialization. As we use $n$ van Emde Boas queues in our algorithm, that would bring the complexity up to $\O{n^2}$, which we do not want to happen.

An alternative (presented in \cite{MehlhornN90}) is to use a hashtable as $H$. The algorithm is now randomized, but every $H$ needs space proportional to the number of stored elements, and initialization is constant-time. The complexity is now $\O{m}$, where $m$ is the total number of elements in $T$, which is all right for our algorithm.

We can, however, de-randomize the algorithm, by simply cutting off several deepest levels of recursion. More precisely, we set a threshold of $K = \log^c{n}$ and we replace every recursive structure $T'$ with size  $ |T'| \leq K^2$ with a standard dictionary on integers (for example, a red-black tree). For the structures $H$ on higher levels we use original $0/1$ arrays. As $|H|$ is always $\O{\sqrt{|T|}}$, the largest $H$ still left has size at most $K$, so the total memory bits used by all $H$ is $n/K$. The dictionaries, on the other hand, use $\O{m}$ space. Every procedure call will now need, apart from standard $\O{\log \log n}$, a single call to a dictionary of size $K$, which works in $\O{\log K} = \O{c \log \log n}$. Therefore, the complexity of a single operation is still $\O{\log \log n}$, with constant factor dependent on $c$. The space needed (and initialization time) is now $\O{m + n/K} = \O{m + \frac{n}{\log^c{n}}}$, as required.

\section{Lower bound for significant pairs}\label{sec:howmanypairs}

In this section we present two sequences $A$ and $B$ with $|A| = |B| = n$, which have at least $\Om{\frac{n^2}{\log n}}$ significant pairs between them. Actually, we will construct, for every integer $k$, two sequences of size $\Th{k \cdot 2^k}$ with $k \cdot 2^{2k}$ significant pairs. To do that, we borrow a construction from \cite{DurajKP17}. In section 3.2 of that paper there is a definition -- for every integer $k$ -- of two integer sequences $A_k, B_k$, each being a concatenation of $2^k$ blocks $\alpha_k^{i}$ or $\beta_k^j$:

$A_k = \alpha_k^0 \circ \alpha_k^1 \circ \ldots \circ \alpha_k^{2^k-1}$

$B_k = \beta_k^0 \circ \beta_k^1 \circ \ldots \circ\beta_k^{2^k-1}$

where $\circ$ denotes concatenation of sequences. To be more precise, we define $A_0 = B_0 = (1)$ and $A_{k+1}, B_{k+1}$ recursively from $A_k, B_k$ as follows: for $i\in\{0,1,\ldots,2^k-1\}$,

\begin{align*}
\alpha_{k+1}^{2i} &= \mathop{inflate}(\alpha_k^i) \circ (2s_k + 2), & 
\beta_{k+1}^{2i} &= \mathop{inflate}(\beta_k^i) \circ (2s_k + 1),  \\
\alpha_{k+1}^{2i+1} &= (2s_k+1, 2s_k+3), &
\beta_{k+1}^{2i+1} &= (2s_k+2, 2s_k+3). \\
A_{k+1} &= \alpha_{k+1}^{0} \circ \alpha_{k+1}^{1} \circ \ldots \circ \alpha_{k+1}^{2^{k+1}-1} &
B_{k+1} &= \beta_{k+1}^{0} \circ \beta_{k+1}^{1} \circ \ldots \circ \beta_{k+1}^{2^{k+1}-1}\\
\end{align*}

where $s_k$ denotes the largest integer of $A_k$ and $B_k$, and the $\mathop{inflate}$ operation ,,doubles'' the sequences in the following way: 
\[ \mathop{inflate}(x_1, x_2, \ldots, x_t) = (2x_1, 2x_1+1, 2x_2, 2x_2+1, \ldots, 2x_t, 2x_t+1) \]

It is then proved that $|A_k| = |B_k| = \Th{k \cdot 2^k}$ and for every $i, j \in \{0, 1, \ldots, 2^k-1\}$,  $\lcis{|\alpha_k^0\ldots\alpha_k^i|, |\beta_k^0\ldots\beta_k^j|} = i + j + 2^k$. 

To adapt this construction to our needs, we introduce $k$ new symbols $\tau_1, \ldots, \tau_k$, each one larger than all current elements of $A_k$ and $B_k$, with $\tau_1  < \tau_2 < \ldots < \tau_k$, and add $(\tau_k, \tau_{k-1}, \ldots, \tau_1)$ at the end of each block. Formally, define for $i = 0, 1, \ldots, 2^k-1$:
\begin{align*}
\alpha'{}_{k}^j = \alpha_{k}^j \circ (\tau_k, \tau_{k-1}, \ldots, \tau_1) \\
\beta'{}_{k}^j = \beta_{k}^j \circ (\tau_k, \tau_{k-1}, \ldots, \tau_1) \\
A'{}_k = \alpha'{}_k^0 \circ \alpha'{}_k^1 \circ \ldots \circ \alpha'{}_k^{2^k-1} \\
B'{}_k = \beta'{}_k^0 \circ \beta'{}_k^1 \circ \ldots \circ\beta'{}_k^{2^k-1}.
\end{align*}

It is clear that $|A'_k| = |B'_k| = \Th{k \cdot 2^k}$, as we have added exactly $k \cdot 2^k$ new symbols to each sequence. Pick $r$ between $1$ and $k$ -- we will now prove that all $\tau_r$-pairs are significant. In each block $\alpha'{}^k_i$ there is a unique occurrence of $\tau_r$ -- let us call its index in $A'_i$ by $x_i$, and similarly, let $y_j$ be the index of the unique occurrence of $\tau_r$ in the block $\beta'{}^k_j$. We will now show that:
\[ \lcisto{x_i, y_j} = i + j + 2^k + 1 \]

\noindent by easy induction on $r$ -- this is enough for all $\tau_r$ pairs to be significant. 

Indeed, we already know from the previous construction that the length of LCIS between $i$-th and $j$-th blocks without any new symbols would be $i+j+2^k$. We can then obtain a common subsequence of length $i+j+2^k+1$ by adding $(x_i, y_j)$. It remains to show that no longer LCIS is possible. If there was one, it would have to contain more than one new symbol -- at least one $\tau_s$-pair for some $s < r$ -- immediately before $(x_i,y_j)$. But if $(x^*_i, y^*_j) = \prv{x_i,y_j}$ is a $\tau_s$-pair for some $\tau_{s} < \tau_r$, then these $\tau_s$ symbols cannot come from the blocks $i$ and $j$ (as $\tau_s$ appears after $\tau_r$), so they come from some blocks $i^* \leq i-1$ and $j^* \leq j-1$. By induction hypothesis, $\lcisto{x^*_i, y^*_j}  \leq (i-1) + (j-1) + 2^k + 1$, so this hypothetical LCIS cannot be longer than $i + j + 2^k + 1$.

Therefore  every matching $\tau_r$-pair is significant, for $r = 1, 2, \ldots, k$. This accounts for $k \cdot 2^{2k}$ significant pairs between $A'_k$ and $B'_k$.

\end{document}